\documentclass[journal]{vgtc}                     

\onlineid{1049}

\vgtccategory{Research}

\vgtcpapertype{please specify}

\title{Enhancing Patient Acceptance of Robotic Ultrasound \\through Conversational Virtual Agent and Immersive Visualizations}

\author{%
  \authororcid{Tianyu Song}{0000-0002-8428-9651},
  Felix Pabst, 
  \authororcid{Ulrich Eck}{0000-0002-5322-4724}, and 
  \authororcid{Nassir Navab}{0000-0002-6032-5611}
}

\authorfooter{
  \item
  	Tianyu Song, Felix Pabst, Ulrich Eck and Nassir Navab are with Technical University of Munich.
  	
  	E-mail: \{tianyu.song, felix.pabst, ulrich.eck, nassir.navab\}@tum.de.

}

\abstract{%
  Robotic ultrasound systems have the potential to improve medical diagnostics, but patient acceptance remains a key challenge. 
To address this, we propose a novel system that combines an AI-based virtual agent, powered by a large language model (LLM), with three mixed reality visualizations aimed at enhancing patient comfort and trust. The LLM enables the virtual assistant to engage in natural, conversational dialogue with patients, answering questions in any format and offering real-time reassurance, creating a more intelligent and reliable interaction. The virtual assistant is animated as controlling the ultrasound probe, giving the impression that the robot is guided by the assistant.
The first visualization employs augmented reality (AR), allowing patients to see the real world and the robot with the virtual avatar superimposed. 
The second visualization is \replaced{a virtual reality (VR) environment with a passthrough window}{an augmented virtuality (AV) environment}, where the real-world body part being scanned is visible, while a 3D Gaussian Splatting reconstruction of the room, excluding the robot, forms the virtual environment. 
The third is a fully immersive virtual reality (VR) experience, featuring the same 3D reconstruction but entirely virtual, where the patient sees a virtual representation of their body being scanned in a robot-free environment. In this case, the virtual ultrasound probe, mirrors the movement of the probe controlled by the robot, creating a synchronized experience as it touches and moves over the patient’s virtual body. 
We conducted a comprehensive agent-guided robotic ultrasound study with all participants, comparing these visualizations against a standard robotic ultrasound procedure. Results showed significant improvements in patient trust, acceptance, and comfort. Based on these findings, we offer insights into designing future mixed reality visualizations and virtual agents to further enhance patient comfort and acceptance in autonomous medical procedures.
}


\teaser{
  \centering
  \includegraphics[width=\linewidth]{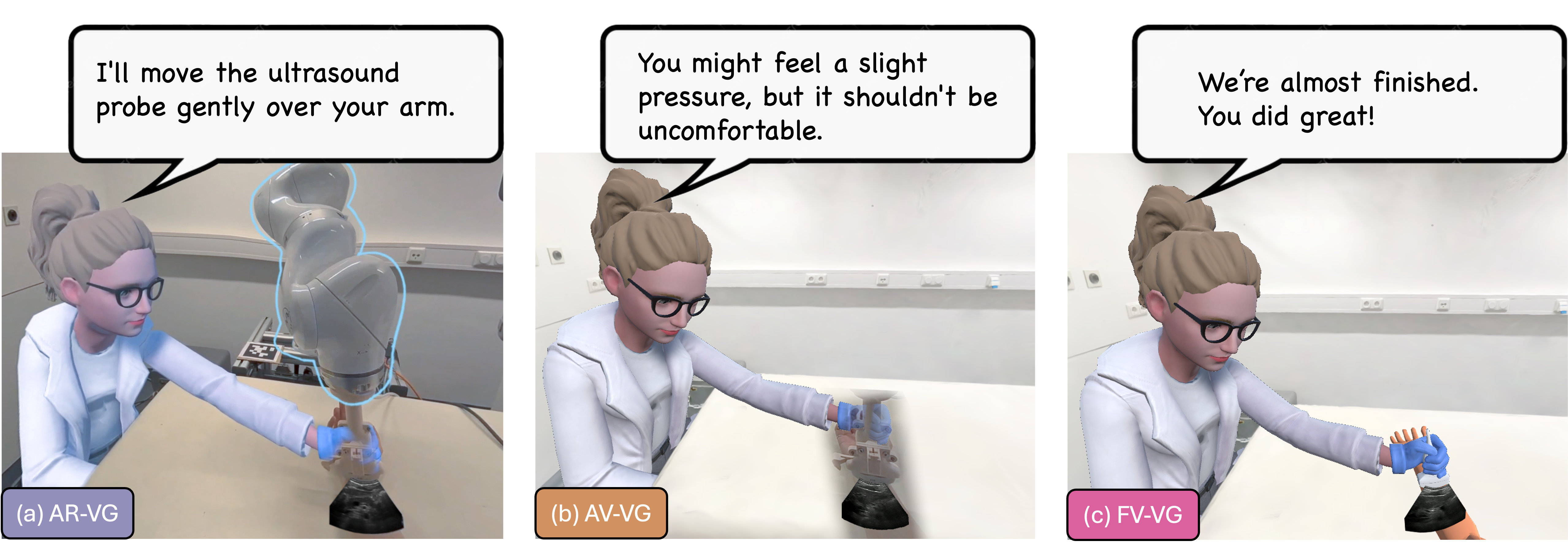}
  \caption{\textbf{Proposed Visualizations with a Conversational Virtual Agent Guiding the Robotic Ultrasound Procedure.} The figure shows three immersive visualizations as seen through the head-mounted display (from left to right): augmented reality, \replaced{virtual reality with a passthrough window}{augmented virtuality}, and fully immersive virtual reality. In both \add{augmented virtuality and }virtual reality visualizations, a realistic 3D Gaussian Splatting model of the actual room, excluding the robot, was used as the virtual environment. In each mode, the virtual human assistant communicates with the patient, offering explanations and reassurance to reduce stress and improve the patient’s experience during the autonomous ultrasound procedure.}
  \label{fig:teaser}
}

\graphicspath{{figs/}{figures/}{pictures/}{images/}{./}} 

\usepackage{tabu}                      
\usepackage{booktabs}                  
\usepackage{lipsum}                    
\usepackage{mwe}                       

\usepackage{mathptmx}                  
\usepackage{soul}
\newcommand{\revised}[1]{\textcolor{black}{#1}}
\newcommand{\add}[1]{\textcolor{black}{#1}}

\newcommand{\replaced}[2]{%
  \textcolor{black}{#2}%
}

\begin{document}


\firstsection{Introduction}

\maketitle

\label{sec:Introduction}

Trust and acceptance are critical for the widespread adoption of autonomous systems, whether in healthcare~\cite{alqudah2021technology,liu2022roles} or in other fields~\cite{choi2015investigating,parsonage2023trust}. \revised{In contexts such as self-driving cars, where automation is designed to minimize human error and enhance safety, public hesitation often arises from a lack of trust~\cite{liu2019evaluating}. }This reluctance often stems from a lack of trust—people are uncomfortable with the idea of handing over something as personal and potentially dangerous as driving to a machine, even if the machine is proven to be highly effective. In healthcare, this issue is even more sensitive, as patients’ physical well-being and emotional comfort are directly impacted by their interactions with robotic systems. Studies show that patients’ hesitation towards robotic systems stems from concerns about trust, privacy, and ethical issues, which directly affect their behavioral intentions toward accepting these technologies in healthcare settings~\cite{kamani2023patients}. Particularly in procedures where human touch and empathy traditionally play a large role, without acceptance, robotic systems may be met with hesitation or rejection by patients~\cite{xu2018would}. \revised{In particularly robotic ultrasound diagnosis, where the patient is conscious throughout the procedure, these concerns become even more pronounced. Robotic ultrasound systems offer advantages in terms of precision and repeatability~\cite{jiang2023robotic,bi2024machine}, but their acceptance has lagged due to the disconnect patients feel when interacting with machines instead of human operators~\cite{eilers2023importance}. This discomfort is magnified by the fact that patients undergoing ultrasound procedures are accustomed to direct human interaction—something robots inherently lack. Therefore, the challenge in designing robotic medical systems is not merely one of technical accuracy but also of creating an experience that is more human-centered, one that addresses patient comfort, trust, and acceptance.}


One \revised{promising solution to} bridge between robotic efficiency and human empathy is the integration of virtual agents that can interact with patients in real-time, offering the reassurance that many patients need in clinical settings~\cite{lisetti2015now}. \add{Recent studies highlight that virtual agents not only improve user engagement but also play a crucial role in fostering trust, particularly when tasks are delegated to these agents~\cite{sun2024digital}. \revised{These agents create a} trust-based interaction framework, which combines rational, emotional, and technological dimensions, underscores the importance of designing agents that balance efficiency with empathy, thereby bridging the gap between automation and human-centric care.} Such agents, when combined with immersive technologies like augmented reality (AR) and virtual reality (VR), can create a hybrid experience where the precision of robotics is augmented by the comforting presence of an AI-driven virtual assistant. This idea forms the core of our approach\add{, aiming to make robotic medical systems more human-centered and patient-friendly}.

In this paper, we present a unique solution aimed at enhancing patient acceptance of robotic ultrasound procedures by combining an AI-based virtual human assistant with immersive visualizations. The virtual assistant interacts with patients throughout the procedure, offering reassurance and explaining the process. It also appears to control the robotic ultrasound probe, giving the impression of human-like involvement. By simulating human presence, we aim to bridge the gap between robotic automation and patient comfort.

To achieve this, we build upon proposed robotic ultrasound system and introduce three immersive visualizations, each designed to enhance the patient experience during the robotic ultrasound. The first is an AR visualization, where the patient sees the real world with the virtual assistant superimposed in the environment, as if present in the room (see Fig.~\ref{fig:teaser}a). The second is \replaced{a VR visualization with a passthrough window}{an Augmented Virtuality (AV) visualization}, which shows the body part being scanned in the real world while the rest of the environment is represented by a 3D Gaussian Splatting model of the room, excluding the robot (see Fig.~\ref{fig:teaser}b). Finally, the fully immersive VR visualization creates a complete virtual environment where the patient views a virtual version of their own body being scanned and synchronized with the real-world procedure (see Fig.~\ref{fig:teaser}c).
These immersive visualizations are designed to address the core issue of patient acceptance by providing a more humanized interaction during robotic ultrasound procedures and enhancing the overall experience.
The key contributions of this paper are as follows:

\begin{itemize}
  \item A novel pipeline that integrates a virtual human assistant and immersive visualizations into robotic ultrasound procedures, providing a structured approach to enhancing patient interaction and comfort.
  \item Three novel visualization modalities designed to enhance patient acceptance of robotic ultrasound systems.
  \item A user study in which we evaluate the impact of these visualizations on patient acceptance and mental workload, demonstrating the significant benefits of our approach.
  \item Insights for future design of mixed reality visualizations and virtual agents to enhance patient comfort and acceptance in autonomous medical procedures.
\end{itemize}

\add{Building on these contributions, the study evaluates three hypotheses that explore how the proposed conversational virtual agents and immersive visualizations address stress, trust, comfort, and usability challenges in robotic ultrasound systems, aligning with the goal of enhancing patient-centered care.} In addition, we make the code publicly available~\footnote{\url{https://github.com/stytim/Robotic-US-with-Virtual-Agent}} to encourage further research and development in enhancing patient acceptance of robotic medical systems.
\section{Related Work}\label{sec:RelatedWork}

\subsection{Trust and Acceptance in Robotic Procedures}

Robotic procedures have become increasingly prevalent in modern healthcare from minimally invasive surgeries~\cite{mack2001minimally,vitiello2012emerging} to autonomous diagnostic tasks~\cite{roshan2022robotic,jiang2023robotic}. However, the integration of these technologies introduces a new set of challenges, particularly in the areas of trust and acceptance, which are critical for both surgeons and patients.
Traditionally, the robot-surgeon relationship has been the primary focus of research in robotic-assisted procedures~\cite{mcdermott2020gender,sierra2021expectations,vichitkraivin2021factors}. Surgeons are often required to trust the accuracy and responsiveness of robotic systems to perform high-stakes operations. Studies have shown that trust between surgeons and robotic systems is built on factors such as system reliability, ease of use, and predictability during procedures~\cite{devito2010clinical,benmessaoud2011facilitators,chatterjee2024advancements,patel2024technical}. However, as robotic systems become more autonomous, the robot-patient relationship is gaining increasing attention~\cite{szabo2024robots}. Acceptance plays an even more critical role when robots interact with fully conscious patients. Studies from Bodenhagen et al.~\cite{bodenhagen2017influence} and Fischer et al.~\cite{fischer2018increasing} showed that transparency enhances trust, which can be increased through clear communication between the robot and patient. Weigelin et al.~\cite{weigelin2018trust} suggested to use verbal cues in addition to kinesthetic interactions to foster trust more effectively. For robotic ultrasound system, Eilers et al.~\cite{eilers2023importance} indicated that pre-examination interactions can lower patient stress levels and significantly enhance the patient acceptance. However, their study did not incorporate AR or VR visualizations, which could further impact patient comfort and trust. 

In addition to direct interactions with the robot, trust can also be affected by other factors. \replaced{Adams et al.\cite{adams2003trust} identified multiple influences on trust in automated systems, including the properties of the automated system, as well as the user’s propensity to trust and the context in which the system is used.}{Adams et al. \cite{adams2003trust} identified three key dimensions affecting trust in automated systems: properties of the system itself, the user’s propensity to trust, and the context of system use.} \replaced{Similarly, Hancock et al.~\cite{hancock2011meta} proposed a triadic model of trust that classifies the factors influencing trust into three categories: human, robot, and environmental characteristics.}{Expanding on this, Hancock et al. \cite{hancock2011meta} proposed a triadic model of trust, categorizing influencing factors into human, robot, and environmental characteristics.} Although robot characteristics, especially performance-based factors, were found to have the greatest impact on perceived trust in Human-Robot Interaction (HRI), environmental factors also moderately affect trust. By incorporating AR and VR environments during robotic procedures, we aim to explore how these additional contextual factors affect trust and acceptance in robotic medical procedures. This is the first work to integrate immersive visualizations in a robotic ultrasound system, providing real-time feedback and interaction to enhance patient trust, comfort, and overall acceptance.

\subsection{Virtual Agents}

Virtual agents, which are digital entities designed to simulate human-like interactions, have become a vital interface for bridging human-machine communication~\cite{cassell2001embodied,nass2000machines}. The embodiment of virtual agents, particularly their visual appearance and behavior, is crucial in fostering user engagement. Ring et al.~\cite{ring2014right} suggest that design rules for an agent's appearance vary by the application domain. Participants found the cartoon character (e.g. enlarged head) friendlier, but the human-proportioned character more appropriate for medical discussions. Studies by Latoschik et al.~\cite{latoschik2017effect} and Zibrek et al.~\cite{zibrek2018effect} further confirmed the benefits as realistic embodiments enhance users’ subjective experiences and increase immersion.

The effectiveness of virtual agents also extends beyond appearance. The integration of verbal and non-verbal behaviors, such as speech, gaze, and gestures, is critical to making interactions feel natural and engaging. As Cowell et al.~\cite{cowell2005manipulation} found, agents displaying nonverbal cues like eye contact and blinking elicit higher levels of trust compared to those lacking these behaviors. Potdevin et al.~\cite{potdevin2021virtual} highlighted how Embodied Conversational Agents (ECA) improve user engagement and intimacy by using voice and animated interactions, which are more effective than text-based methods. Kopp et al.~\cite{kopp2008modeling} emphasized the importance of integrating multimodal behaviors in ECAs. They demonstrated how these behaviors, alongside verbal interactions, create a more cohesive and lifelike communication experience, making the agent more relatable and engaging for users. 
 
 In healthcare applications, virtual agents play an increasingly important role. Nadarzynski et al.~\cite{nadarzynski2019acceptability} explored the acceptability of AI-led chatbot systems for healthcare and came to the conclusion that due to absence of empathy and a professional human touch made the chatbots to some users less acceptable. This highlights the possibility of extending a chatbot with a human-like virtual agent to increase the acceptance of the system. Philip et al.~\cite{philip2017virtual} studied the acceptability of an ECA in a face-to-face clinical interview done to diagnose major depressive disorders. Patients rated the interview with the ECA as highly acceptable, indicating that the ECA can convey empathy, build patient trust, lessen feelings of judgment from a human, and lower emotional barriers to sharing their affective state. The system incorporated a speech synthesizer for the ECA and a speech recognizer for the patient's answers but lacks the ability to freely generate the virtual human's responses~\cite{philip2014could}. Lucas et al.~\cite{lucas2017reporting} also demonstrated that virtual human interviewers can enhance service members' disclosure of mental health symptoms. Moreover, virtual agents in mixed reality environments offer unique opportunities for blending virtual and physical interactions~\cite{holz2011mira,norouzi2020systematic}. For example, Kim et al.~\cite{kim2019blowing} demonstrated how subtle environmental interactions, such as airflow influencing both virtual and real objects, can increase the sense of social presence in mixed reality environments by making virtual agents seem more aware of and connected to the physical space around them. This highlights the potential of agents that can perceive and interact with both digital and physical elements in healthcare settings. These agents could further enhance patient trust by offering real-time feedback during complex procedures, such as robotic surgeries or autonomous diagnostic tasks~\cite{juravle2020trust}.
However, as reviewed by Laranjo et al.~\cite{laranjo2018conversational}, the use of conversational agents  with unconstrained natural language input capabilities in healthcare is still in the early stages of investigation. This highlights the importance of advancing the integration of conversational, non-verbal, and empathetic behaviors to make these agents more effective in scenarios where patient comfort, trust, and clear communication are critical. In the context of robotic ultrasound procedures, we believe that virtual agents have the potential to significantly humanize interactions. By offering real-time feedback and guidance, we aim to show they can bridge the gap between automation and human empathy, ultimately improving patient acceptance and overall user experience.



\subsection{Effects of Level of Immersion}

\revised{Milgram's} mixed reality continuum~\cite{milgram1995augmented} introduces a spectrum that ranges from fully real environments to fully virtual environments, with varying levels of augmented and mixed realities in between. Understanding the effects of immersion is crucial for designing systems that optimize user experience, particularly in healthcare, where patient comfort and engagement are paramount.

Past works have analyzed the impacts that different levels of immersion have on the user in a various contexts. Mania et al.~\cite{mania2001effects} compared four different conditions including real, 3D desktop, 3D head mounted display (HMD) and audio-only for a 15-minute seminar presentation, studying how levels of immersion affect memory recall, memory awareness, and perceptions of the experimental space and sense of presence. While higher presence did not always correlate with accurate memory recall, both presence and semantic memory recall were significantly higher in the ``real'' condition. Ragan et al.~\cite{ragan2010effects} compared low and high immersion in a procedure memorization task by differing field of view of the user directly and of the software and the field of regard. They found that higher levels of immersion resulted in better performance, while specifically stating that lower-cost VR systems showed statistically significant performance improvements compared to conditions with lower immersion levels. Pollard et al.~\cite{pollard2020level} implemented three different levels of  immersive technology: a desktop monitor and a static audio source (low level), a partially occlusive, mid-grade HMD with supra-aural headphones (medium level) and a fully occlusive HMD with circumaural headphones (high level). For the task of an ordered scavenger hunt followed by questions about observed objects and their spatial relationships in the environment, the high immersion condition was found to improve object recognition compared to the medium and low level.
Liberatore and Wagner’s systematic review~\cite{liberatore2021virtual} further supports the importance of immersion by aggregating results from multiple empirical studies, showing that fully immersive VR is effective in creating stress-relief environments or in treating phobias, while AR tends to be more useful in situations where a mix of virtual and real elements are needed, such as rehabilitation tasks.
Drawing from previous studies, we aim to understand how these levels of immersion could affect patient trust, comfort, and acceptance during robotic procedures.

\section{Methods}\label{sec:Methods}

\begin{figure*}
    \centering
    \includegraphics[width=\textwidth]{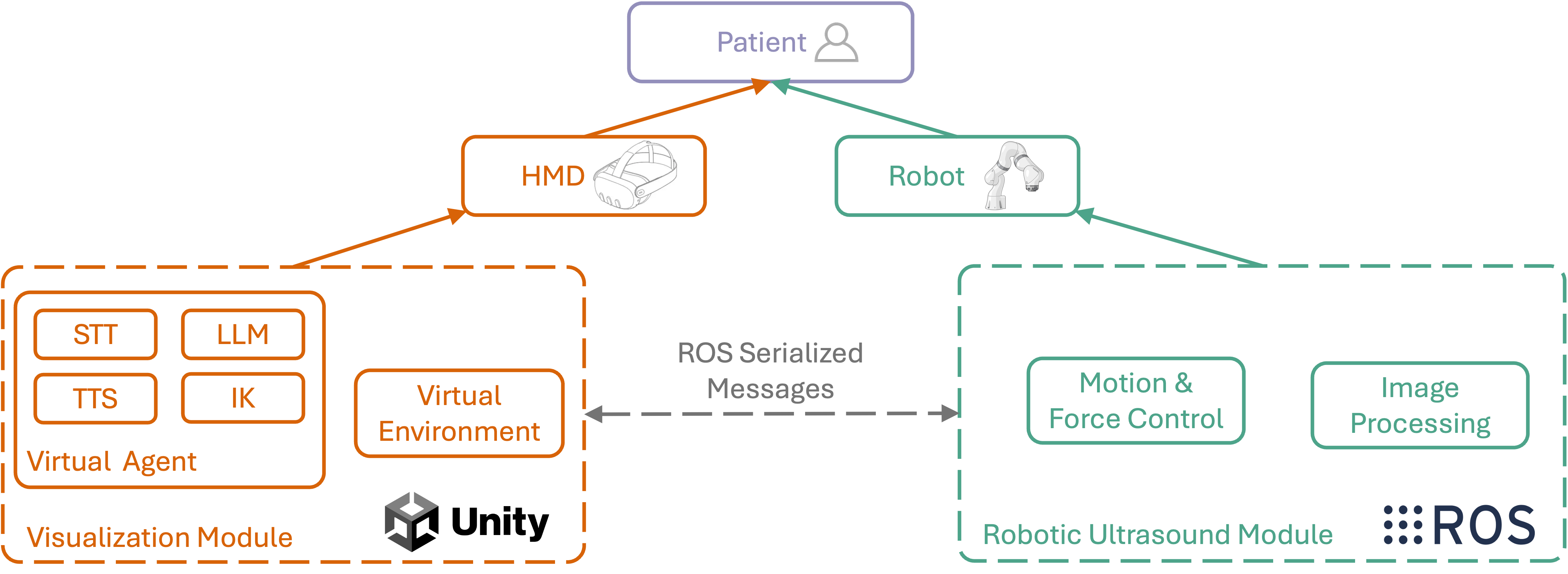}
    \caption{\textbf{Overview of the Proposed Pipeline.} The pipeline adds a visualization module, including augmented reality and virtual reality modalities, enhanced by a virtual human assistant. The system architecture facilitates real-time interaction and synchronization between the robotic ultrasound system, the virtual agent, and the patient, providing a more engaging and patient-centered experience.}
    \label{fig:overview}
\end{figure*}

In this work, we propose a pipeline, as illustrated in Fig.~\ref{fig:overview}, that introduces a visualization module into the existing robotic ultrasound system (RUS), enhancing the patient experience through conversational virtual agent and immersive visualizations. 
In the following subsections, we detail the components and theoretical frameworks behind the conversational virtual agent, the different visualization modalities (AR, \revised{AV}, and fully immersive VR), as well as the registration process that ensures the seamless integration of virtual and real-world elements.

\subsection{Force-Compliant Robotic Ultrasound System}

Our robotic ultrasound system employs a force-compliant control approach based on impedance control~\cite{jiang2020automatic,jiang2021autonomous,jiang2021deformation}, which ensures safe and accurate probe positioning during the ultrasound procedure by regulating both contact force and probe orientation. This control system operates as a spring-like mechanism with predefined stiffness values, ensuring that if an obstacle, such as the patient’s body, is encountered, the robot will either bypass with a reduced force or stop at the location to avoid applying excessive force.

Impedance control modulates the interaction forces between the robotic arm and the patient’s tissue by adjusting the robot’s stiffness, damping, and inertia parameters. The control law is defined as:
\begin{equation}
\tau = J^T \left[ F_d + K_m e + D \dot{e} + M \ddot{e} \right]
\end{equation}
where $\tau$ represents the computed joint torques, $J^T$ is the transposed Jacobian matrix, and $e$ is the pose error between the current and target positions. The stiffness, damping, and inertia matrices are denoted by $K_m$, $D$, and $M$, respectively. The desired force $F_d$ controls the applied contact force. To ensure stable force along the probe’s centerline, the robot uses a 1-DOF compliant controller for force control and a 5-DOF position controller for precise positioning.

This force-compliant control \revised{follows the same formulation used in Hennersperger et al.\cite{hennersperger2017towards} and Jiang et al.\cite{jiang2021autonomous}, which} allows the system to adapt to soft tissue deformations and variations in probe position, preventing excessive force that could cause discomfort or harm. \add{To enable communication between the RUS and the visualization module, messages are serialized and passed between the two. This ensures that the visualization module is always aware of the robot’s state and the current phase of the procedure. Additionally, commands produced through user interactions with the conversational virtual agent in the visualization module can be sent back to the RUS via serialized messages. This bidirectional communication enables real-time synchronization and seamless integration of robotic control and immersive visualizations.}

\subsection{Conversational Virtual Agent}

The conversational virtual agent serves as a central component of the proposed visualizations, aiming to enhance patient interaction by natural communication and reduce the sense of isolation commonly associated with robotic procedures~\cite{almeras2019operating}, therefore creating a more humanized and engaging experience.

At the core of the agent’s functionality is a speech-to-text (STT) system that transcribes the patient’s spoken words into textual input. The transcribed text is then processed by a large language model (LLM), which generates contextually appropriate responses based on pre-configured prompt. The use of an LLM ensures that the agent provides responses that are both relevant to the patient’s situation and emotionally supportive, fostering a sense of trust and ease during the procedure.
Once the response is generated, it is converted back into audible speech through a text-to-speech (TTS) engine, which provides natural output. The TTS system matches the assistant’s appearance and personality, ensuring consistent and realistic verbal interactions. \add{Similar technical pipelines of the conversational intelligent agent have been used in domains such as virtual museums~\cite{garcia2024speaking} and educational VR applications~\cite{yang2024effects}.}

In addition to the agent’s verbal responses, various animations are employed to enhance its realism and human-like presence~\cite{yu2021avatars}. These include eye blinking, mouth movements synchronized with speech generated by the TTS system, and a subtle idle breathing animation to give the avatar a more natural appearance. Beyond these baseline animations, inverse kinematics (IK) is used to further enhance realism by controlling the assistant’s head and hand movement. The IK algorithm calculates natural, human-like gestures in response to the patient’s position and the state of the robot. \revised{Studies have shown that non-verbal cues such as gestures, head movements, and subtle body language significantly enhance user immersion and realism in virtual environments~\cite{etienne2023perception, beck2012emotional}. Adapting these findings to our robotic ultrasound use case, we implemented a set of non-verbal behaviors to increase the agent’s presence and engagement.} For example, the assistant’s head turns toward the patient when they speak, and its hand animates to hold the ultrasound probe when the robot moves it within the agent’s arm reach, as though the virtual assistant were guiding the procedure. This visual and behavioral consistency strengthens the impression of the virtual agent being physically present and actively engaged with the patient.

Together, these elements form the foundation of the conversational virtual agent framework. The integration of real-time speech, intelligent language processing, and naturalistic physical interaction ensures that the agent serves as a comforting presence, enhancing the overall patient experience.

\subsection{Augmented Reality Visualization}

The AR visualization, illustrated in Fig.~\ref{fig:teaser}a, is designed to blend virtual elements with the patient’s real-world environment, allowing for seamless integration of the guidance from virtual human agent within a familiar, physical space. This approach maintains situational awareness by allowing the patient to see both their surroundings and the robotic arm at work while interacting with the assistant. 
In this visualization, the avatar appears seated next to the patient, engaging with them through real-time conversation. Through the HMD, the patient can see both the virtual assistant and the ultrasound probe, which the avatar is holding and guiding over the patient’s body. As the robotic arm moves the ultrasound probe in the physical world, the avatar’s virtual hand mimics these motions synchronously. This visual configuration provides the patient with the comforting illusion that the human-like avatar is controlling the robotic procedure, reducing the sense of detachment associated with the automated process.

\subsection{\revised{Augmented Virtuality} Visualization}

The \replaced{VR Passthrough}{AV} visualization, as demonstrated in Fig.~\ref{fig:teaser}b, combines an immersive virtual environment with selective real-world visibility through a passthrough window. This allows the patient to remain visually engaged with the area of their body being scanned while still interacting with the virtual human assistant in the virtual world. The virtual environment mirrors the physical room, but the robotic ultrasound system itself is not visible; only the ultrasound probe becomes visible when it enters the passthrough window, allowing the patient to see it during the scan.
Within the virtual environment, the patient can also see a virtual representation of the ultrasound probe, which mirrors the position of the real probe. As the robotic arm drives the real ultrasound probe toward the patient, the virtual probe moves correspondingly, providing a visual cue that the probe is approaching. When the probe transitions into the passthrough window, the virtual probe seamlessly aligns with and transitions into the real probe, ensuring continuity and reassuring the patient that the procedure is progressing as expected.

In this visualization, the virtual human assistant interacts with the patient in the same way as in the AR setting. The assistant is positioned to give the impression of guiding the robotic ultrasound. The patient’s view of the scan area remains in the passthrough window.
This visualization preserves the patient’s sense of presence and control by allowing them to observe their body during the scan. At the same time, the virtual environment and virtual human assistant help to reduce anxiety with guidance and interaction throughout the procedure.

\subsection{Fully Immersive Virtual Reality Visualization}

The fully immersive VR visualization, as shown in Fig.~\ref{fig:teaser}c, places the patient entirely in a virtual environment, removing any direct visual connection to the real world. In this setting, the virtual environment mirrors the physical room, but the robotic ultrasound system is absent as well, reinforcing the illusion that the virtual assistant is in full control of the procedure. The patient perceives the virtual agent as the sole operator, guiding the ultrasound probe, which enhances the feeling of human involvement and control.

In this visualization, the patient sees a virtual replica of their own body. As the robot moves the ultrasound probe on the patient’s actual body, the virtual probe in the VR environment moves in sync on the virtual body. This synchronization ensures that the patient feels a consistent tactile and visual connection between what they see in the virtual space and what they feel in the real world.
This immersive environment is designed to reduce anxiety by eliminating the often sterile, impersonal atmosphere of traditional robotic procedures, replacing it with a comforting virtual experience where the patient feels human involvement and control throughout. By maintaining real-time synchronization with their physical body and providing a strong sense of presence through the virtual human assistant, the visualization offers a fully immersive alternative that fosters comfort and trust in the procedure.

\subsection{Registration of Virtual and Real Components}

\begin{figure}
    \centering
    \includegraphics[width=0.85\columnwidth]{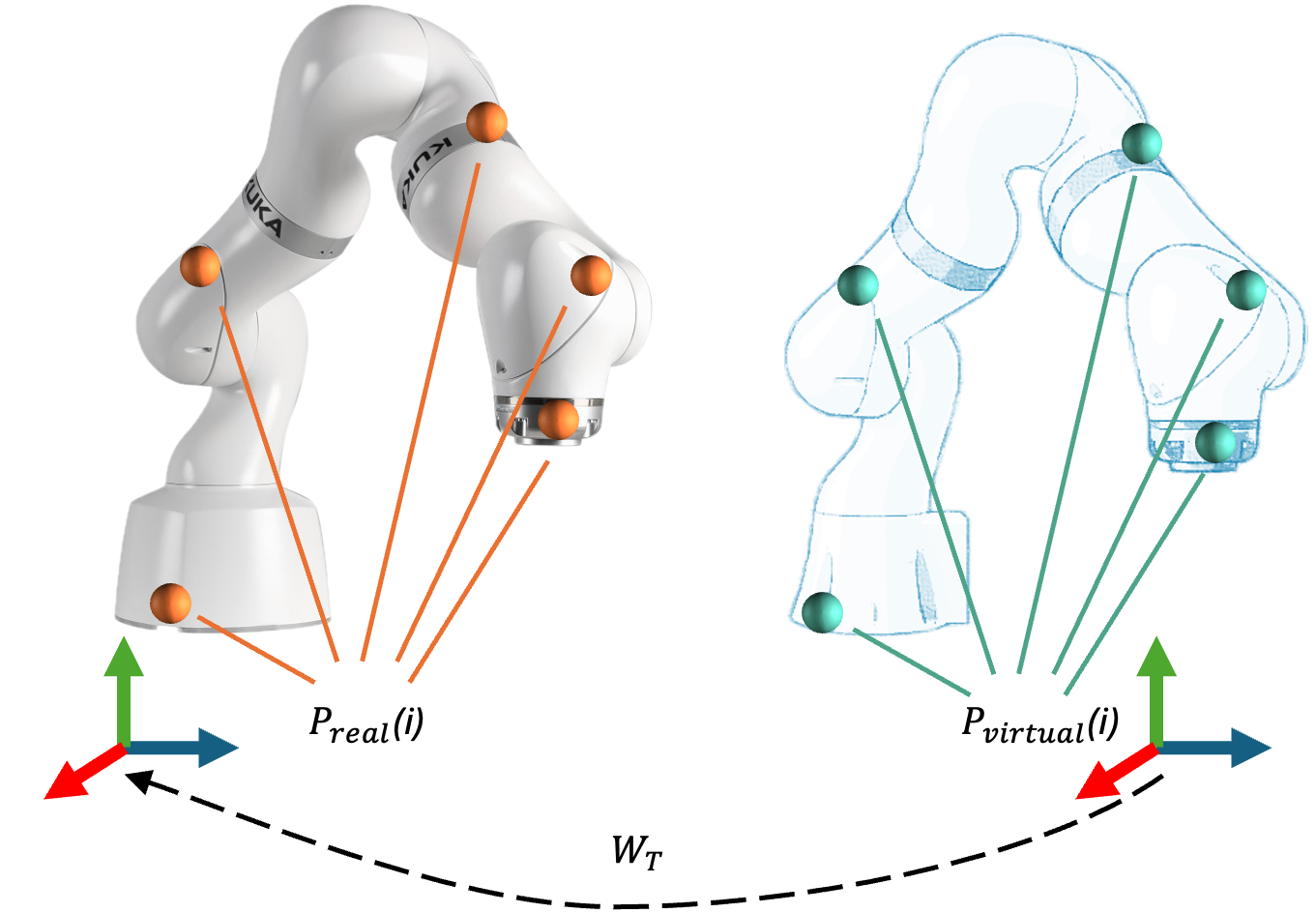}
    \caption{\add{\textbf{Registration of Virtual and Real Robot Using Predefined Points.} Predefined points on the virtual robot are shown in green. We marked the corresponding points on the real robot with the HMD, shown in orange. The dashed lines represent the transformation matrix $W_T$ to be solved, which aligns the two point sets.}}
    \label{fig:registration}
\end{figure}

Accurate registration between the virtual and real-world components is crucial for maintaining spatial coherence across all visualizations. To achieve this, we utilize a set of predefined, ordered points on both the virtual and physical elements to perform 3D-to-3D registration, similar to the approaches used in previous works~\cite{song2022happy,yu2022duplicated}. For instance, several key points $P_{virtual}(i)$ are selected on the virtual robot, which correspond to matching points marked on the real robotic system $P_{real}(i)$. Similarly, for the virtual environment, corner points of the physical room are used to align the virtual representation of the room with the real one. The transformation matrix  $W_T$  between these point sets is computed to minimize the difference between their positions. This can be expressed mathematically as:

\begin{equation}
	W_T = \arg\min_{\widehat{W}_T} d\left(\widehat{W}_T P_{virtual}(i), P_{real}(i)\right)
\end{equation}
where $W_T$ is the estimated transformation matrix between the real and virtual spaces, and $d$ represents the distance between corresponding points in the two spaces. \add{The registration process is visually illustrated in Fig.~\ref{fig:registration}.} We further decompose this transformation into a rotation component $W_R$ and a translation component $W_t$, solving each sequentially using the Kabsch algorithm~\cite{kabsch1993automatic}.

After the initial alignment is achieved, we can leverage the spatial anchor feature offered by modern HMDs to maintain this alignment between sessions. This means the calibration procedure only needs to be performed once, unless fine-tuning or refinements are required in future sessions. By using spatial anchors, the system ensures persistent alignment, enhancing the immersive experience for the patient without the need for repeated calibrations.

\section{User Study}\label{Sec:UserStudy}

To evaluate the impact of conversational virtual agent and different mixed reality visualizations on patient acceptance and comfort during an autonomous robotic ultrasound procedure, we conducted a $1\times4$ within-subject user study. \add{The study was approved by the Ethics Committee of Technical University of Munich under protocol number 2022-87-S-KK.}

\subsection{Hypotheses}

Based on the previous findings of virtual agent and effects of level of immersion, we propose the following three hypotheses:
\begin{itemize}
    \item[\textbf{\textit{H1.}}]\replaced{The presence of a virtual agent will reduce discomfort and increase acceptance. We hypothesize that, compared to the baseline robotic ultrasound procedure (without any visualization), the appearance of a virtual human assistant guiding the procedure will reduce patient discomfort by providing a more humanized and reassuring experience.}{The presence of a virtual agent will improve patient comfort and acceptance. Compared to the baseline robotic ultrasound procedure, incorporating a virtual human assistant during the procedure will enhance the overall patient experience.}
    \item[\textbf{\textit{H2.}}] \replaced{Reduced visibility of the robot will reduce stress and increase trust. We hypothesize that when the robot is not visible to the patient, such as in the \add{AV and }VR environments, the procedure will be perceived as less intimidating compared to conditions where the robot is fully visible.}{The visibility of the robot will influence patient trust and stress levels. The extent to which the robot is visible during the procedure will affect how intimidating the procedure is perceived by the patient, impacting their trust and stress levels.}
	\item[\textbf{\textit{H3.}}] \replaced{The level of immersion will influence patient workload and usability. We hypothesize that AR, \replaced{VR with passthrough}{AV}, and fully immersive VR will have varying effects on usability and mental workload. Specifically, AR will demonstrate better usability due to maintaining better situational awareness, while fully immersive VR may induce higher cognitive.}{The level of immersion will affect patient usability and workload. Different levels of immersion across AR, AV, and VR environments will have varying impacts on usability and mental workload during the procedure.}
\end{itemize}

\subsection{Experimental Variables and Measures}

The study was designed with one independent variable: the mode of visualization applied during the ultrasound procedure. Participants experienced four different conditions. The baseline condition, Robotic Ultrasound (\textbf{RUS}), involved no visualizations. In contrast, the three proposed conditions included immersive visualization modalities: AR Virtual Agent Guidance (\textbf{AR-VG}), \revised{AV} Virtual Agent Guidance (\textbf{\revised{AV}-VG}), and Fully Immersive VR Virtual Agent Guidance (\textbf{FV-VG}).

Several dependent variables were measured to evaluate the effects of these visualization modes. Stress levels were objectively recorded using an ECG sensor, providing physiological data on participant stress before and during the procedure under each condition. In addition, participants’ mental workload was assessed using the NASA Task Load Index (NASA-TLX)~\cite{hart2006nasa}. The user-friendliness and usability of the visualizations was measured using the System Usability Scale (SUS)~\cite{brooke1996sus}. Finally, participants’ trust in the robotic system was evaluated using the HRI Trust Score, a scale originally developed by Schaefer et al.~\cite{schaefer2016measuring} and later modified by Eilers et al.~\cite{eilers2023importance}, to measure the level of trust participants placed in the robotic system during each condition.

\begin{figure}[h]
    \centering
    \includegraphics[width=\columnwidth]{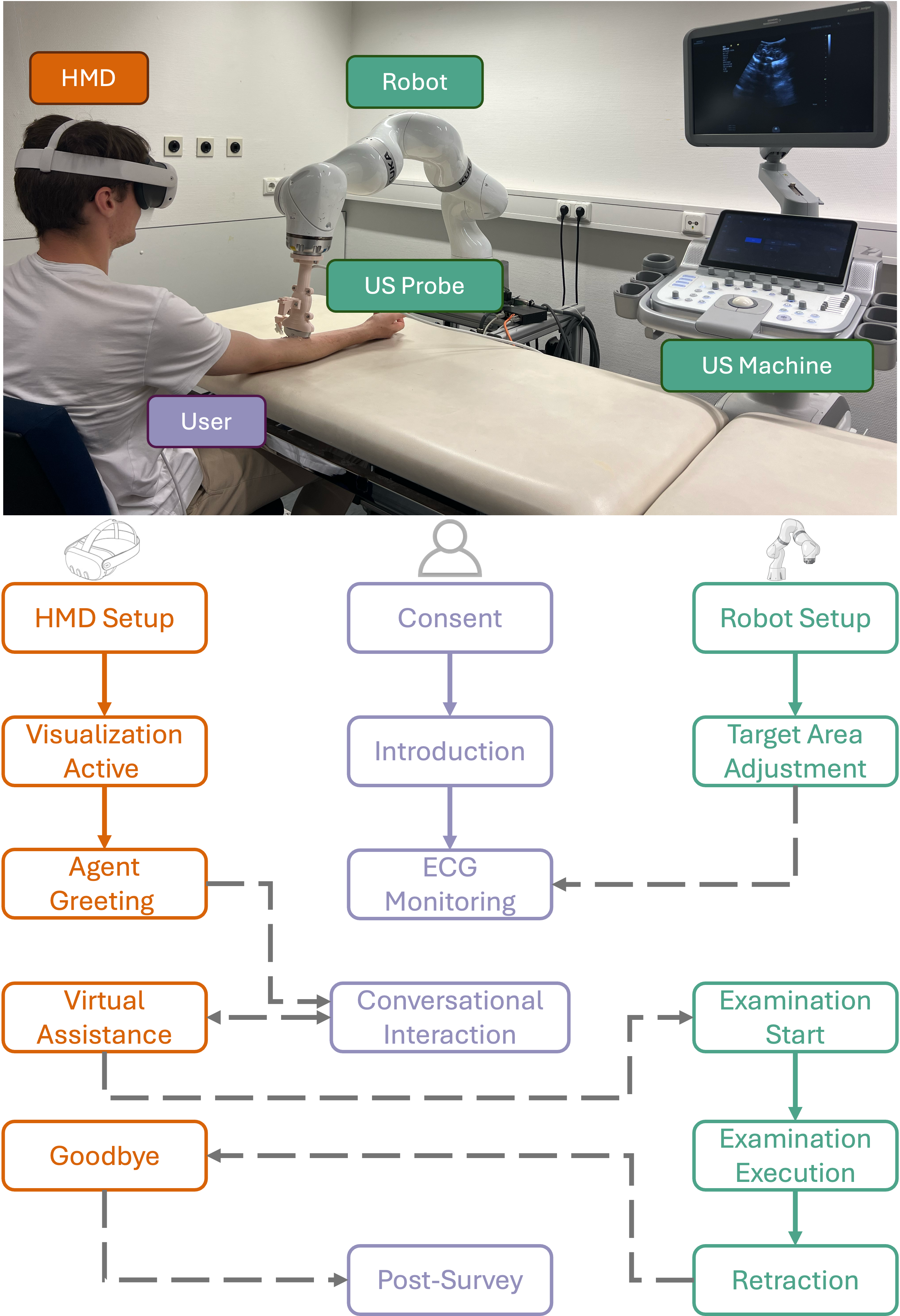}
    \caption{\textbf{User Setup (top) and User Study Flow Chart (bottom)}. The setup includes the patient interacting with the conversational virtual agent and the robotic ultrasound system, using different immersive visualizations provided by the HMD. The flow chart outlines the user study procedure: the orange (left column) indicates the visualization module, the purple (middle column) represents the user, and the green (right column) depicts the robotic ultrasound module. Dashed lines represent the interactions between these three main components.}
    \label{fig:flow}
\end{figure}

\subsection{Setup}

The proposed visualizations are developed using the Unity game engine (Unity Technologies, San Francisco, US), version 2022.3.18f1. Communication between the Unity-based visualizations and the robotic ultrasound system, controlled via the Robot Operating System (ROS), is facilitated through the ROS TCP Endpoint v0.7.0~\cite{unity2022rostcp}. The robotic system is powered by a KUKA LBR IIWA R800 robotic arm (KUKA AG, Augsburg, Germany), which is controlled via IIWA Stack~\cite{hennersperger2017towards} running on ROS Noetic. The ultrasound system used is the Siemens ACUSON Juniper (Siemens Healthineers, Erlangen, Germany), with the transducer mounted to the robotic manipulator using a custom-printed holder. A frame grabber was used to capture the images from the ultrasound machine and stream them directly to the Unity application. Meta Quest 3 (Meta, Menlo Park, US) with firmware version 67.0 is chosen for deployment of the Unity solution for its versatility in supporting varying levels of immersion, which is critical for testing the effectiveness of each visualization approach under the same hardware conditions. Additionally, we utilized the Depth API~\cite{depthapi} to provide more believable visual rendering that handles occlusion between virtual and physical components. 

The virtual human avatar used in the proposed visualizations is created using Ready Player Me~\cite{readyplayerme}. For this study, we designed the avatar as a female figure wearing a doctor’s coat to enhance its professional and comforting presence\add{, as previous research has shown that anthropomorphic features, such as human-like appearance and professional attire, can foster trust and positive attitudes toward avatar-assisted therapy~\cite{moriuchi2023looking}}. The avatar’s animations consist of both baked and procedural elements. Idle animations, such as sitting and breathing, eye blinking, and mouth movement during speech, are baked into the avatar’s model. In contrast, procedural animations—like the avatar reaching to grab the ultrasound probe when it enters its proximity, or turning to face the patient during interaction—are dynamically generated using the inverse kinematics solver Final IK 2.3~\cite{finalik}.
Speech-to-text is implemented via the whisper.unity 1.3.2~\cite{whisperunity}, which runs locally on the system. We selected the Whisper tiny model (OpenAI, San Francisco, US) for its balance between performance and speed, ensuring rapid transcription during patient interaction. The transcribed text is processed using the LLM for Unity 2.2.0~\cite{llmunity}. We opted for the Phi-3 3B model (Microsoft, Redmond, US), which is also run locally, achieving a similar balance between speed and performance to ensure responsiveness during real-time interactions. Once a response is generated by the LLM module, it is fed into the TTS module, implemented using the Meta Quest Voice SDK 67 with Wit.AI for speech synthesis, providing natural and consistent vocal output from the avatar.

To create a virtual environment identical to the real-world setting, we utilized an iPhone 13 Pro equipped with the Luma AI app~\cite{lumaai} to perform a detailed scan of the room. This process generated a 3D Gaussian Splatting model of the environment, capturing the room’s geometry and textures with high accuracy. The resulting model was then imported into Unity, where it was rendered using the Gaussian Splatting VR Viewer~\cite{gaussianvr}. Finally, the visualization system runs on a PC equipped with an NVIDIA RTX 2060 Super GPU, an Intel i7-10700F CPU, and 16GB RAM. The Meta Quest 3 was connected to this PC via Quest Link using a USB-C cable.

\subsection{Participants}

A total of 14 participants (5 female, 9 male) took part in this study, with ages ranging from 23 to 64 years (M = 31.6, SD = 9.9). Participants were required to have no conditions that could hinder the use of AR and VR, although the use of corrective glasses or contact lenses was allowed. \add{Regarding prior experience with AR and VR, 50$\%$ of participants rated themselves below 3 on a 5-point scale, indicating limited experience, while the other 50$\%$ rated themselves 3 or higher.} In terms of familiarity with robotic procedures, 28.6$\%$ of participants reported being not at all familiar to somewhat familiar, 28.6$\%$ were moderately familiar, and 42.8$\%$ were very familiar to expert level.
Participation in the study was entirely voluntary, and participants were free to withdraw at any time. The study adhered to the ethical principles outlined in the Declaration of Helsinki. To ensure confidentiality, all data collected during the study was fully anonymized.

\subsection{Procedure}

The study procedure, as depicted in Fig.~\ref{fig:flow}, began with a registration process prior to the arrival of participants. Using the Meta Quest 3 controller, we placed virtual spheres at predefined points on both the real robot and the room for registration. 
Upon arrival, participants were informed about the study procedure and provided with a consent form outlining the purpose of the study, their role, their rights, and assurances of data confidentiality. Once informed consent was obtained, participants filled out a demographic form. Afterward, they were fitted with a biosignalsplux 3-lead ECG sensor (PLUX Biosignals, Lisbon, Portugal) following a tutorial. The ECG sensor was connected via Bluetooth to a MacBook Pro running OpenSignals software to record physiological data.
Participants were seated in a chair and instructed to place their right arm on an ultrasound exam table. An experienced operator applied ultrasound gel to the participant's arm, and they were asked to keep it still while the robot, equipped with the ultrasound probe, was manually positioned by the operator to record the start and end points of the scanning path. The robot arm was then lifted away from the participants to prepare for the procedure.

Once preparations were complete, ECG recording began, and the visualizations were activated. The virtual avatar greeted the participants, prompting them to engage in conversation. Participants were free to speak with the avatar, raise concerns, or ask questions. They could also instruct the avatar to begin the procedure at their discretion. During the ultrasound procedure, which was performed with the robot arm set to apply the force of 8 N in the Z direction with stiffness of 500 N/m, participants were encouraged to continue interacting with the avatar if desired.
Upon completion, the virtual assistant informed the participant that the procedure was finished, and they could relax and move their arm. The visualization was then deactivated, and the ECG sensor recording was stopped. The presentation order of the different visualization methods—baseline (\textbf{RUS}), \textbf{AR-VG}, \textbf{\revised{AV}-VG}, and \textbf{FV-VG}—was randomized for each participant to avoid bias.
After each task, participants completed the SUS, NASA-TLX, and the HRI Trust Score questionnaire. Upon finishing all tasks for both the baseline and the proposed visualizations, participants were asked to rank the visualizations based on their preferences and provide feedback. In total, each user study session lasted between 30 to 40 minutes.

\subsection{\add{Statistical Analysis}}

\add{To analyze the data collected during the study, we employed a range of statistical tests to evaluate differences across the visualization conditions. 
For stress levels derived from ECG-based heart rate variability (HRV), non-parametric tests were used: Wilcoxon Signed-Rank test for within-condition comparisons (e.g., resting vs. execution phases), and the Kruskal-Wallis test for between-condition comparisons. These methods were selected due to the non-normal distribution of the data as confirmed by the Shapiro-Wilk test.
For subjective measures, a Friedman test was conducted to determine if there were significant differences among the four visualization methods. Post-hoc Dunn-Sid{\'a}k pairwise comparisons were performed to identify specific group differences when the Friedman test indicated significance. Additionally, effect sizes were calculated for pairwise comparisons using Cohen’s $d$ and $\eta^2$ for the Kruskal-Wallis test to quantify the magnitude of observed differences.
All statistical analyses were performed using Python and appropriate libraries, with significance levels set at $\alpha = 0.05$.}

\section{Results}\label{Sec:Results}

The system’s performance was evaluated across key metrics, including latency, frame rate, and resolution. Latency was measured for the key components of the virtual human assistant interaction: STT exhibited a latency of $46 \pm 5$ ms, the LLM processing took $552 \pm 187$ ms, and the TTS synthesis had a latency of $1281 \pm 188$ ms.
The visual output was rendered at a resolution of $4128 \times 2208$ on the HMD, with frame rate recorded to assess the visual fluidity of each visualization modality. The \textbf{AR-VG} visualization maintained a consistent average frame rate of 72 FPS. Both \textbf{\revised{AV}-VG} and \textbf{FV-VG} operated at an average frame rate of 36 FPS.

\subsection{Stress Level}

\begin{figure}
    \centering
    \includegraphics[width=\columnwidth]{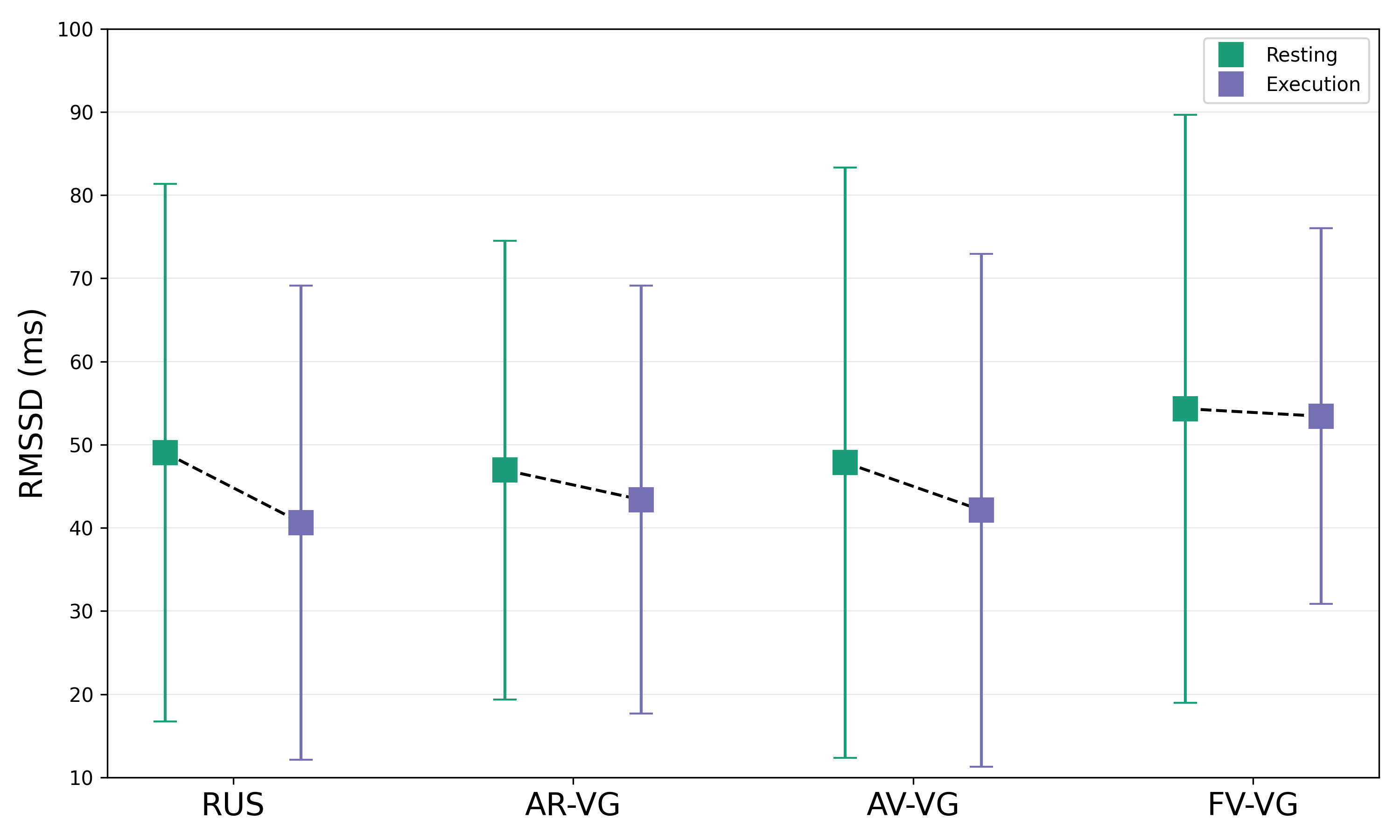}
    \caption{\textbf{HRV during the Resting and Execution Phases.} In \textbf{RUS}, the RMSSD shows the steepest drop between the two phases, indicating a higher stress level compared to the proposed visualizations. \textbf{AR-VG} and \textbf{\revised{AV}-VG} perform similarly, while \textbf{FV-VG} exhibits highest RMSSD value and the smallest change between phases, suggesting that less stress is induced during the execution.}
    \label{fig:hrv}
\end{figure}

To assess stress levels during the robotic ultrasound procedure, we derived HRV from the ECG sensor data, focusing on the Root Mean Square of the Successive Differences (RMSSD), a commonly used measure of stress~\cite{shaffer2017overview}. Lower RMSSD values generally indicate higher stress levels. The analysis was performed using the HeartPy~\cite{van2019heartpy} Python package.
We analyzed HRV during two phases of the procedure: the resting phase, where the robot remained stationary and participants were free to interact with the virtual agent, and the execution phase, during which the robot performed the ultrasound scan. The HRV data for these phases are shown in Fig.~\ref{fig:hrv}.

Given the non-normal distribution of the data observed by the Shapiro-Wilk test, we used the Wilcoxon Signed-Rank Test for within-condition comparisons, assessing differences between the resting and execution phases for each visualization method. Although we observed a trend of lower RMSSD values during the execution phase compared to the resting phase, in \textbf{RUS} ($z = 25.0, p = 0.846\add{, d = 0.291}$), \textbf{AR-VG} ($z = 13.0, p = 0.547\add{, d = 0.141}$), \textbf{\revised{AV}-VG} ($z = 38.0, p = 0.970\add{, d = 0.180}$), and \textbf{FV-VG} ($z = 28.0, p = 0.700\add{, d = 0.032}$), the results did not indicate significance.
To compare HRV across the different visualization methods during both the resting and execution phases, we employed the Kruskal-Wallis Test. The analysis for the resting phase showed no significant difference in HRV across the visualization methods ($H = 0.485, p = 0.922\add{, \eta^2 = 0.012}$). During the execution phase, the test also yielded no significant difference between methods ($H = 3.430, p = 0.330\add{, \eta^2 = 0.086}$).

\subsection{Subjective Ratings}

\begin{figure*}[t]
    \centering
    \begin{subfigure}[b]{0.325\textwidth}
        \centering
        \includegraphics[width=\textwidth]{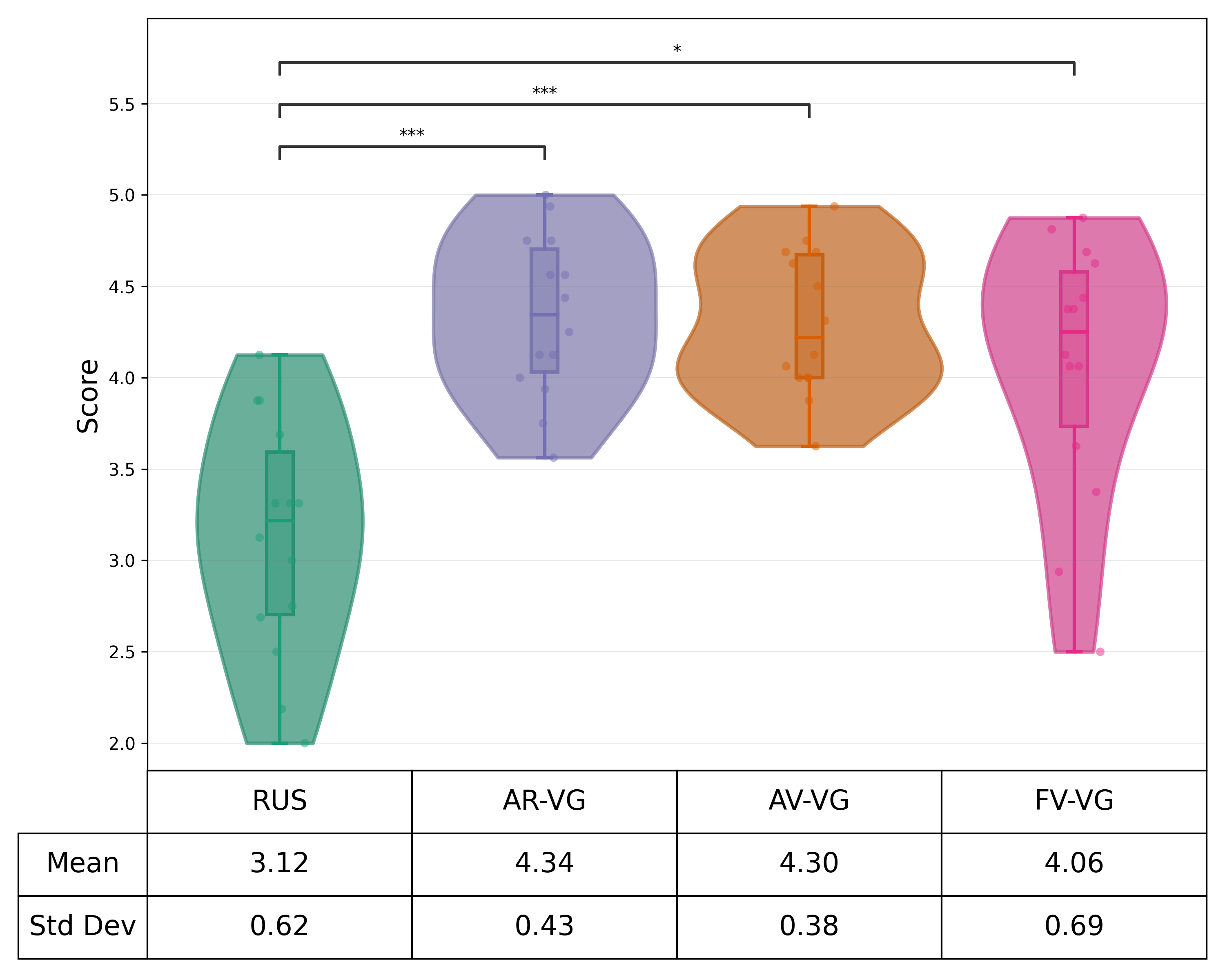}
        \caption{Trust in Human Robot Interaction}
        \label{fig:hri}
    \end{subfigure}
    \hfill 
    \begin{subfigure}[b]{0.325\textwidth}
        \centering
        \includegraphics[width=\textwidth]{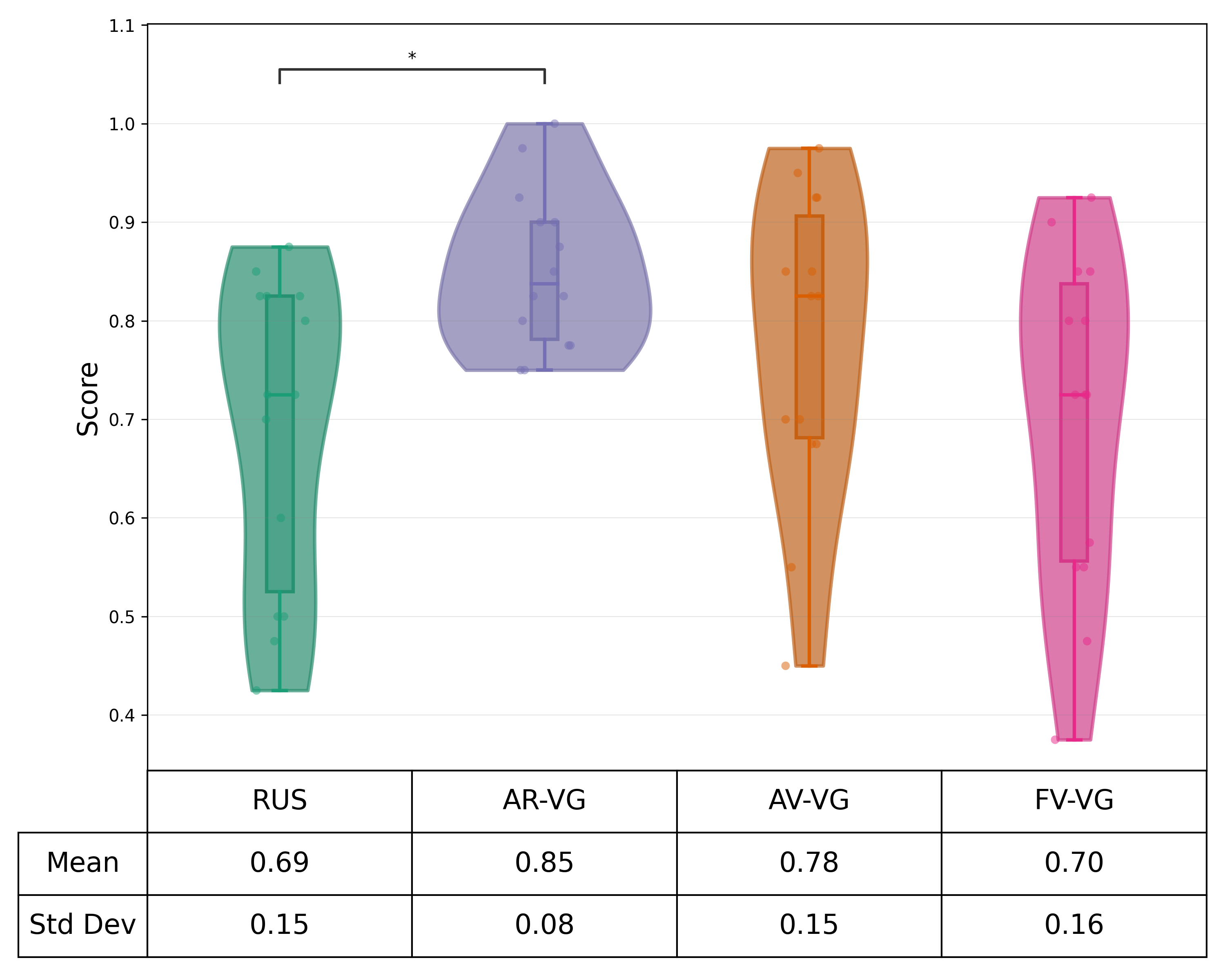}
        \caption{System Usability Score}
        \label{fig:sus}
    \end{subfigure}
    \hfill 
    \begin{subfigure}[b]{0.325\textwidth}
        \centering
        \includegraphics[width=\textwidth]{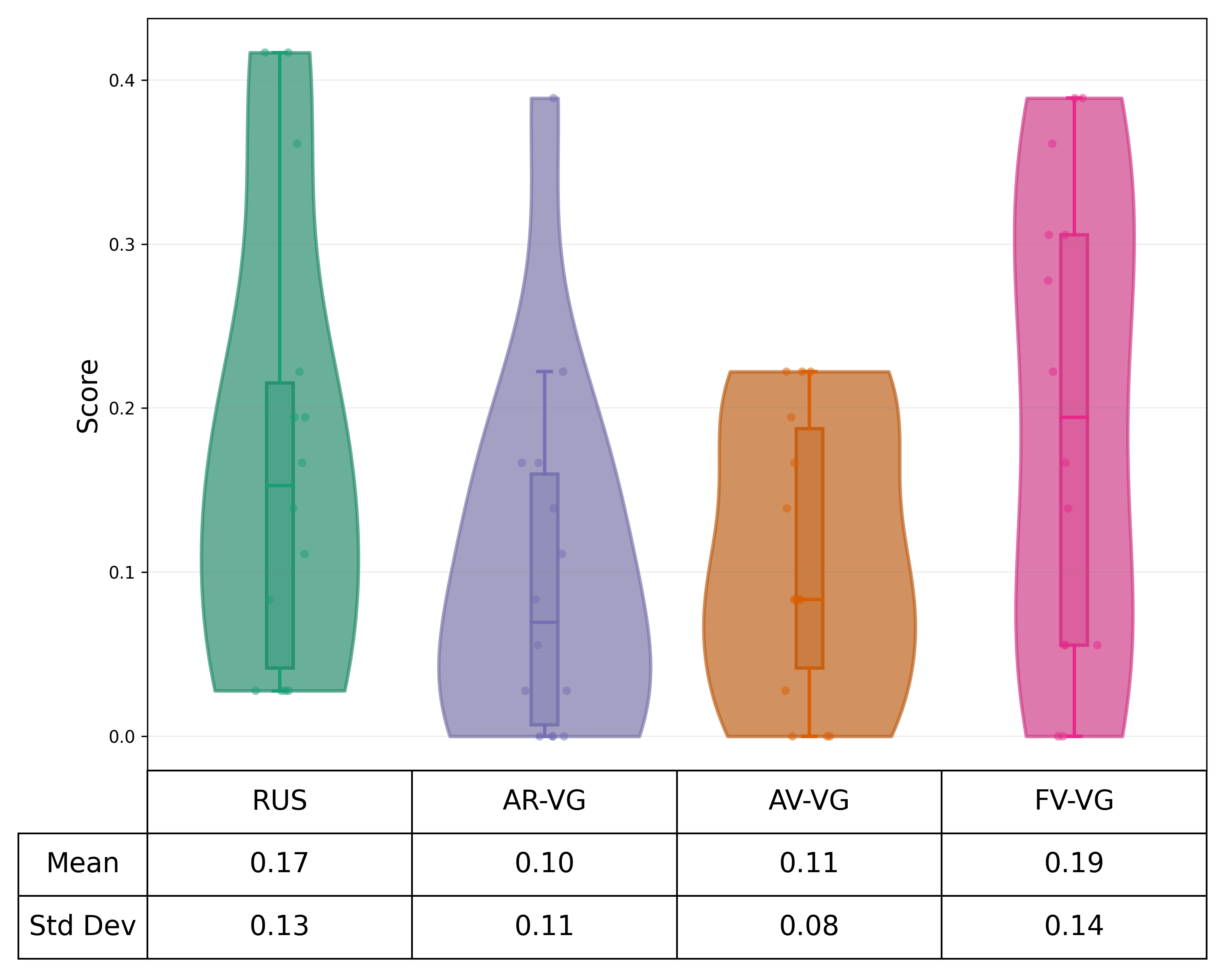}
        \caption{Perceived Workload}
        \label{fig:tlx}
    \end{subfigure}
    \caption{\textbf{Subjective Measurements for Trust Score, Usability, and Workload.} All proposed immersive visualizations with the conversational agent significantly increase the HRI trust score compared to \textbf{RUS}. \textbf{AR-VG} receives the highest trust score, the best usability, and the lowest workload among all methods. Statistical significance is indicated as $\star \left( p<0.05 \right)$, $\star \star \left( p<0.01 \right)$, and $\star \star \star \left( p<0.001 \right)$.}
    \label{fig:subjective}
\end{figure*}

HRI Trust scores under each condition for the robotic ultrasound were as follows: \textbf{RUS} ($M = 3.12, SD = 0.62$), \textbf{AR-VG} ($M = 4.33, SD = 0.42$), \textbf{\revised{AV}-VG} ($M = 4.29, SD = 0.38$), and \textbf{FV-VG} ($M = 4.06, SD = 0.68$). The results are visualized in Fig.~\ref{fig:hri}.
Statistical analysis using the Friedman test revealed a significant difference in trust scores across the visualization methods ($\chi^2(3) = 26.95, p = 6.02 \times 10^{-6}$). Post-hoc Dunn-Sid{\'a}k  pairwise comparisons further emphasized these differences. Significant differences were observed between \textbf{RUS} and \textbf{AR-VG} ($p = 0.000316\add{, d = 2.272}$), \textbf{RUS} and \textbf{\revised{AV}-VG} ($p = 0.00035\add{, d = 2.272}$), and \textbf{RUS} and \textbf{FV-VG} ($p = 0.012\add{, d = 1.428}$).

The SUS scores for each condition\revised{, normalized to a 0-1 scale,} are shown in Fig.~\ref{fig:sus}. A Friedman test revealed a significant difference in usability across the visualization methods ($\chi^2(3) = 16.60, p = 0.000854$). Post-hoc Dunn-Sid{\'a}k  pairwise comparisons indicated that the significant difference lies between \textbf{RUS} and \textbf{AR-VG} ($p = 0.037\add{, d = 1.343}$).

The NASA-TLX scores, \add{normalized to a 0-1 range}, are presented in Fig.~\ref{fig:tlx}. A Friedman test revealed a significant difference in task load across the visualization methods($\chi^2(3) = 9.03, p = 0.02$).
Although there was a tendency for both \textbf{AR-VG} and \textbf{\revised{AV}-VG} to show lower task load scores compared to \textbf{RUS}, Dunn-Sid{\'a}k  pairwise comparisons did not reveal any statistically significant differences between the visualization methods.

\subsection{User Preference and Feedback}


The results showed that \textbf{AR-VG} was the most preferred visualization, with 72$\%$ of participants ranking it as their top choice, 14$\%$ ranking it second, and 14$\%$ ranking it third. \textbf{\revised{AV}-VG} followed, with 21$\%$ of participants ranking it as the most preferred, 43$\%$ ranking it second, and 36$\%$ ranking it third. For \textbf{FV-VG}, 36$\%$ of participants ranked it in their top three choices. Finally, no participants ranking \textbf{RUS} as their first choice. However, 22$\%$ ranked it second, 42$\%$ ranked it third, and 36$\%$ ranked it as their least preferred visualization.

The qualitative feedback from participants provided further insight into their preferences. Participants in general appreciated the conversational abilities of the virtual assistant across \textbf{AR-VG}, \textbf{\revised{AV}-VG} and \textbf{FV-VG}. They noted that talking to the avatar felt natural and gave them more control over the procedure. In addition, several participants remarked that the hand animation of the virtual assistant taking control of the probe “made me trust the system more.” However, due to technical limitation, the avatar’s hand was not visible in the \textbf{\revised{AV}-VG} passthrough window, which led to some confusion about the interaction.
Concerns about the accuracy of VR visualizations were also raised. Participants noted that due to tracking error, sometimes misalignment between their real and virtual arms in \textbf{FV-VG} caused uncertainty about the success of the scan. Participants raised concerns about the robot’s actions, particularly when they could not see the real robot.

Overall, the feedback indicated that participants favored the visualizations that offered a balance between immersion and real-world visibility and integrating a friendly, responsive avatar can improve patient trust and comfort in robotic ultrasound procedures.

\section{Discussion}\label{Sec:GeneralDiscussion}

\subsection{Hypotheses}
\textbf{\textit{H1.}} Our results demonstrate a significant increase in trust scores across all the proposed visualizations featuring the conversational virtual agent, compared to \textbf{RUS}. This finding strongly supports the hypothesis that the presence of the virtual agent contributes to reducing discomfort and increasing acceptance during the robotic ultrasound procedure. Participants were able to ask questions, receive feedback, and feel reassured by the agent’s presence, which appears to have played a key role in fostering trust. Notably, several participants highlighted the hand animation of the virtual assistant holding the ultrasound probe, describing it as a crucial factor in building trust. This subtle yet meaningful interaction gave participants the impression that the virtual agent was aware of the ongoing procedure, making the system appear more intelligent and responsive. By simulating the action of guiding the probe, the virtual assistant conveyed a sense of human control, reducing the perceived detachment often associated with autonomous systems. This visual synchronization between the agent’s actions and the real-world procedure helped humanize the experience, further enhancing confidence in the system’s accuracy and reliability.
Moreover, the usability of the system also improved across all the proposed methods featuring the virtual agent, although significant improvements in usability were only observed with \textbf{AR-VG}. The perceived workload was also reduced in both \textbf{AR-VG} and \textbf{\revised{AV}-VG}. The agent’s conversational abilities, particularly in offering explanations and responding to patient inputs, likely reduced the cognitive burden and made the system easier to navigate.

\textbf{\textit{H2.}} Our results provide partial support for the hypothesis that reducing the visibility of the robot will reduce stress and improve acceptance. When comparing the three immersive visualization methods, \textbf{FV-VG} showed the highest RMSSD values among all conditions in both the resting and execution phases, with the smallest change between these phases. This suggests that participants experienced the least increase in stress during the procedure in the fully immersive environment, potentially due to the absence of the robot’s visual presence, which could reduce feelings of intimidation or anxiety. \add{However, the lack of statistical significance across conditions indicates that the visualization methods may primarily influence psychological perceptions—such as reduced anxiety and improved comfort—rather than inducing measurable changes in physiological stress responses. Additionally, a larger sample size may increase the power of statistical analyses and reveal trends not observed in this study.}
Participant feedback also highlighted the varied reactions to the lack of robot visibility. One participant with no prior experience in robotic procedures noted that in \textbf{FV-VG}, the environment felt like “an animated world,” allowing them to focus less on the procedure itself. This participant expressed a sense of relief and detachment from the robotic aspect, commenting that “before you realize it, the procedure is done.” This suggests that for those unfamiliar with robotic systems, full immersion may help reduce anxiety by removing any focus on the technical aspects of the procedure.
However, several participants with more experience in robotics, especially those with development experience, expressed discomfort with not being able to see the robot’s movements. These participants indicated they would prefer to observe the robot, as they were concerned about the possibility of malfunction or errors. This feedback aligns with the lower trust scores for \textbf{FV-VG}, compared to \textbf{AR-VG} and \textbf{\revised{AV}-VG}, despite the reduced physiological stress. The misalignment between the real and virtual bodies in \textbf{FV-VG}, combined with the complete absence of visual cues from the robot, likely contributed to a lower sense of control and trust in the system.
In contrast, \textbf{AR-VG}, where the robot is visible alongside the virtual agent, had the highest trust scores. This suggests that for many participants, being able to observe the robot’s actions provided reassurance and increased their trust in the system. Similarly, \textbf{\revised{AV}-VG}, where the robot was hidden but the patient’s real arm was visible, performed well in terms of trust, though slightly lower than \textbf{AR-VG}. These findings indicate that while reducing the robot’s visibility may lower stress, maintaining some visual connection to the real world, whether through the robot or the patient’s body, is crucial for building trust.

\textbf{\textit{H3.}} Our results indicate support for the hypothesis that the level of immersion influences patient workload and usability. Among the three immersive visualization methods, \textbf{AR-VG} demonstrated the highest usability and the lowest perceived workload. This supports the hypothesis that AR, by maintaining a connection to the real world, allows for greater situational awareness, which makes the system easier to navigate and reduces cognitive effort. Participants could see their surroundings and the virtual agent, making the experience more intuitive and less mentally taxing. The blend of real-world context with virtual elements likely contributed to both the higher usability and the lower workload.
\textbf{\revised{AV}-VG} also performed well in terms of both usability and workload, though slightly below \textbf{AR-VG}. The passthrough window, which allowed participants to see their real arm during the procedure, offered a partial connection to the real world while still immersing them in a virtual environment. This balance between immersion and real-world visibility may have helped reduce mental load compared to \textbf{FV-VG}, as participants were reassured by seeing part of their real body. However, the higher level of immersion compared to \textbf{AR-VG} might have slightly increased cognitive effort, resulting in a moderate workload and usability score.
In contrast, \textbf{FV-VG} demonstrated the lowest usability and the highest perceived workload among the three methods. The fully immersive environment removed all real-world visual cues, requiring participants to rely entirely on the virtual environment and the virtual agent for orientation and guidance. This complete detachment from the real world may have contributed to a sense of disorientation, which in turn negatively impacted usability and increased increased cognitive demand, as participants had to adapt to the fully virtual setting.

\subsection{Insights}

\textbf{Context-aware Communication.}
The importance of context-aware communication from the virtual agent was a key finding in this study, and it aligns with broader research in human-robot interaction~\cite{chevalier2022context}. In medical settings, patients often feel anxious or disconnected from autonomous systems due to the perceived lack of transparency and control. By embedding a conversational agent that is aware of the procedure’s stages—beginning, execution, and ending—our system ensured that patients were continuously informed and reassured. This type of communication reduces uncertainty, which is crucial in maintaining trust and comfort, as seen in other works that emphasize the role of transparency in fostering trust in autonomous systems~\cite{ososky2014determinants,pynadath2018transparency}. Context-aware systems that adjust feedback based on the current state of the procedure, as we implemented, align with research suggesting that timely, relevant communication enhances user experience and trust~\cite{lisetti2015now}. Moreover, while automating feedback can reduce patient cognitive load, it is important to avoid over-automation, as excessive automation can lead to a loss of sense of agency (SoA)~\cite{haggard2012sense,ueda2021influence} and potentially increase stress, especially in medical contexts where patient involvement is critical.

\textbf{Balancing Immersion and Real-World Context.}
One of the key insights from our study is the delicate balance between immersion and real-world context in patient experience during robotic ultrasound. While participants generally preferred \textbf{AR-VG} and \textbf{\revised{AV}-VG}, the stress levels were actually lower in \textbf{FV-VG}. This suggests that while a highly immersive environment can reduce physiological stress, it may also disconnect patients from critical real-world cues, such as the robot’s actions, which are crucial for maintaining trust and confidence. Research has shown that users tend to feel more comfortable when they have some level of real-world feedback, particularly in medical settings, where understanding the procedure is important for reducing anxiety~\cite{burghardt2018non,weisfeld2021dealing}. To address this, a potential future design could combine the benefits of both approaches. For instance, in a fully immersive VR environment, or even a calm, relaxing virtual setting, abstract representations of the robot’s state could be introduced. This would allow patients to enjoy the calming benefits of the VR environment while still being aware of the robot’s movements, thus providing both stress reduction and a sense of control. Such a hybrid visualization approach could balance immersion with real-world awareness, enhancing both comfort and trust in autonomous medical procedures.

\textbf{Patient-Centered Design.}
\add{This study represents a first step toward integrating conversational virtual agents and immersive visualizations into robotic ultrasound systems.}
A key takeaway from this study is the importance of designing immersive visualizations with the patient’s experience at the forefront, particularly in procedures where patients remain conscious. Our findings emphasize that immersive technologies should not merely serve as technical enhancements but must also be tailored to meet the emotional and psychological needs of patients. The inclusion of a conversational virtual agent, for example, not only humanized the procedure but also helped reduce feelings of isolation and discomfort by providing constant reassurance.
A patient-centered approach can extend beyond medical robotics to other fields where human interaction with autonomous systems is critical. For example, future designs should prioritize personalization~\cite{athanasiou2014towards}, allowing systems to adapt to individual patient preferences, whether through adjusting levels of immersion, offering more or less transparency during the procedure, or tailoring communication styles to the patient’s comfort level. Additionally, systems can be designed to remember previous interactions, enabling the virtual agent to build rapport by referencing past experiences. For instance, when a patient returns for a follow-up visit, the system could greet them and mention something from the previous session, helping to create a more familiar and personalized interaction. Ultimately, this approach ensures that patients remain active participants in their own care, which is essential for fostering long-term trust and acceptance of autonomous technologies.

\subsection{Limitations}
\revised{While the study demonstrates promising results as a proof-of-concept, several limitations and trade-offs should be addressed in future work.}

First, tracking inaccuracies in \textbf{FV-VG} affected user confidence, with some participants reporting misalignment between their real and virtual bodies. This issue stems from two factors: 1) inaccurate hand tracking from the HMD, and 2) the IK solver estimating the arm pose based solely on the hand and head positions. To mitigate this, adding additional sensors to the arm could improve tracking accuracy. However, this would increase the complexity of the setup, which could negatively impact usability. \add{Additionally, these inaccuracies may have introduced biases, placing \textbf{FV-VG} at a disadvantage compared to other conditions. Caution is warranted when interpreting its results, as differences may stem from technical issues rather than the visualization method itself. Future studies should refine tracking mechanisms to ensure fair comparisons and address this imbalance.}

Second, while participants appreciated the conversational abilities of the virtual agent, some reported delays in communication, leading to uncertainty about whether their input was received. To address this, incorporating a visual indicator, such as the avatar nodding its head, or audio feedback, like the avatar quickly responding with a verbal acknowledgment such as “uh-hum,” could help reassure users that their input has been recognized. Additionally, the LLM powering the virtual agent could be further enhanced by training it on more specific data. This would enable the agent to provide more professional and accurate answers during interactions, improving the overall user experience.

Third, in the \textbf{\revised{AV}-VG} implementation, due to technical limitations, virtual elements are not visible in the passthrough window. This is because the Unity Meta Quest SDK only allows the virtual layer to be rendered either above or below the real-world layer, but not mixed. This impacts usability and overall experience, as participants felt less confident without being able to see the avatar holding the ultrasound probe. Exploring other headsets or custom rendering engines that offer more flexibility in how virtual and real-world content is layered might provide a better user experience.

Furthermore, participants noted a depth perception issue in both AR and VR visualizations. The ultrasound image attached to the probe always appeared on top of the patient’s arm, causing patients to misjudge the probe’s position. This misjudgement led to doubts about the system’s accuracy. In the future, improving the visualization of the ultrasound image—such as by adjusting its transparency when it intersects with the arm—could help resolve this issue and provide a more realistic and reassuring experience.
\add{Finally, the participant pool was skewed toward participants with prior knowledge of robotics platforms. While this demographic provided valuable insights into the usability and technical aspects of the system, it may not fully represent the target population—patients with limited exposure to robotic systems. This may have influenced the trust and acceptance measures observed in the study. Additionally, participants’ acceptance of wearing an HMD during the procedure was not separately measured but was included in overall acceptance and trust ratings for the visualizations. Another limitation is the inability to separate the effects of the virtual avatar from those of the ultrasound probe visualization. For example, trust may have increased due to the avatar, the probe visualization, or their combined effects, while usability in \textbf{VP-VG} may have been impacted by physical-virtual alignment errors. Future studies should recruit a more diverse participant pool and explore alternative methods of delivering immersive visualizations to ensure broader applicability.}
\section{Conclusion}\label{Conclusion}

In this paper, we presented a novel system aimed at enhancing patient acceptance of robotic ultrasound procedures through the integration of immersive mixed reality visualizations and an AI-based conversational virtual agent. Our system was evaluated across three different visualization modalities—\textbf{AR-VG}, \textbf{\revised{AV}-VG}, and \textbf{FV-VG}—each designed to improve patient trust, comfort, and usability compared to baseline robotic ultrasound.
The results of our study demonstrated that the inclusion of a conversational virtual agent significantly increased patient trust and reduced discomfort, with \textbf{AR-VG} emerging as the most preferred visualization method. While fully immersive VR reduced physiological stress, the participants expressed a stronger preference for visualizations that maintained some connection to the real world, as seen in \textbf{AR-VG} and \textbf{\revised{AV}-VG}. This balance between immersion and real-world context appears to be critical for maintaining both comfort and trust.
Overall, our findings provide valuable insights into how virtual agents and mixed reality visualizations can be leveraged to improve patient comfort and trust in autonomous medical procedures. By combining technological advancements with a focus on patient-centered design, future systems can further bridge the gap between automation and human empathy, paving the way for broader acceptance of robotic medical devices.

\acknowledgments{%
	This work was partly supported by the state of Bavaria through Bayerische Forschungsstiftung (BFS) under Grant AZ-1592-23-ForNeRo. The authors would like to thank all NARVIS and IFL Lab members for their valuable help and feedback.%
}

\bibliographystyle{abbrv-doi-hyperref}

\bibliography{09_Bibliography}

\end{document}